\documentstyle[aps,multicol,pra,epsfig]{revtex}
\begin{document}

\draft

\title{Variational approach to the modulational instability}
\author{Z. Rapti$^1$, P.G. Kevrekidis$^1$, A. Smerzi$^{2,3}$ and 
A.R. Bishop$^3$}
\address{
$^1$  Department of Mathematics and Statistics, University of 
Massachusetts, Amherst, MA 01003-4515, USA \\
$^2$ Istituto Nazionale di Fisica per la Materia BEC-CRS and
Dipartimento di Fisica, Universita' di Trento, I-38050 Povo, Italy\\
$^3$ Theoretical Division and Center for Nonlinear Studies, Los Alamos
National Laboratory, Los Alamos, NM 87545, USA\\
}
\date{\today}
\maketitle
\begin{abstract}
We study the modulational stability of the nonlinear Schr\"odinger 
equation (NLS) using a time-dependent variational approach. 
Within this framework, we derive ordinary 
differential equations (ODEs) for the time evolution of the amplitude and
phase of modulational perturbations. Analyzing the ensuing ODEs, we re-derive
the classical modulational instability criterion. The
case (relevant to applications in optics and Bose-Einstein condensation)
where the coefficients of the equation are time-dependent, is also examined.
%Within this framework,
%the stability of the system can be studied in terms of a 
%simple mechanical analog, namely with a classical particle 
%moving in an effective potential.
%A further consequence of our approach is that we can easily study the
%quantum (many-body) version of the NLS quantizing only the degrees
%of freedom of the effective particle. We conclude that, in contrast
% with the classical case, the quantum system is always
%modulationally unstable.
\end{abstract}

%\pacs{PACS: 63.20.Pw, 05.45.-a}
%\begin{multicols}{2}
\vspace{2mm}

Modulational instability (MI) is a general feature of
discrete as well as continuum nonlinear wave equations.
This instability shows that in such settings, a specific
range of wavenumbers of plane wave profiles of the form
$u(x,t) \sim \exp(i (k x - \omega t))$ becomes unstable
to modulations. The latter effect leads to an exponential 
growth of the unstable modes and eventually to delocalization
(upon excitation of such wavenumbers) in momentum space. 
That is, in turn, equivalent to localization in position space,
and hence the formation of localized, coherent solitary wave structures
\cite{sulem}.

The realizations of this instability 
span a diverse set of disciplines ranging
from fluid dynamics \cite{benjamin67} (where it is usually referred
to as the Benjamin-Feir instability) 
and nonlinear optics \cite{ostrovskii69}
to plasma physics \cite{taniuti68}.
One of the earliest contexts in which its significance
was appreciated was the linear stability analysis 
of deep water waves. 
It was much later recognized that the
conditions for MI would be significantly modified for discrete 
settings relevant to, for instance,
the local denaturation of DNA \cite{peyrard93} 
or coupled  arrays of optical waveguides 
\cite{morandotti99}. 
In the latter case,
the relevant model is the discrete nonlinear Schr\"odinger equation (DNLS), 
and its MI conditions were discussed in \cite{kivshar92}. 
Most recently, the MI has been recognized as responsible for dephasing
and localization
phenomena in the context of Bose-Einstein condensates (BEC) in the
presence of an optical lattice \cite{wu01,sm02,konotop02,cat02}.

In this brief report, we present an alternative approach to
the modulational stability of plane waves in the context of the 
nonlinear Schr{\"o}dinger equation
\begin{eqnarray}
i \psi_t=-\psi_{xx}- U |\psi|^2 \psi,
\label{seq1}
\end{eqnarray}
where $\psi$ is a complex field, the subscripts denote partial
derivatives with respect to the corresponding variable and $U$
is a constant prefactor (the strength of the nonlinearity).
We examine MI using a time dependent variational approach (TDVA), 
and study the results
in comparison with the standard linear stability (LS)
calculations.
It should be mentioned that the use of TDVA for the
study of solitons at the classical \cite{malomed} and even
at the quantum \cite{treppiedi} level is not novel. 
What distinguishes our study from these earlier ones is the
use of the MI-motivated ansatz in the TDVA (see below). We also
note in passing that MI and solitons from a quantum mechanical
point of view have been considered in a number of references 
including (but not limited to) \cite{pra}. 
Furthermore,
a similar in spirit, 3-mode approximation was systematically
developed in the works of \cite{trillo}. There are however a number
of differences between the latter and the present approach such
as e.g., our use of the variational formulation of the problem
(instead of the application of the 3-mode ansatz in the dynamical
equation in the context of \cite{trillo}), as well as the fact
that we are perturbing around an exact plane wave solution, while
in the case of \cite{trillo}, the plane wave is an additional mode
in the relevant expansion.

In the linear stability framework (see e.g., \cite{kivshar92,sm02} and
references therein for relevant details), 
the stability of the plane waves has been examined. The latter are of
the form
\begin{eqnarray}
\psi(x,t)=\psi_0 \exp \left[i \left(k x - \omega t \right) \right],
\label{seq2}
\end{eqnarray}
and constitute exact solutions of the nonlinear Schr{\"o}dinger equation
with a dispersion relation 
\begin{eqnarray}
\omega=k^2 - U \psi_0^2.
\label{seq3}
\end{eqnarray}
Then, the MI is examined in the LS framework using the linearization
\begin{equation}
u(x,t)=(\psi_0+\epsilon c) \exp[i ((k x-\omega t) +\epsilon d(x,t))]
\label{seq4}
\end{equation}
and analyzing the $O(\epsilon)$ terms as
\begin{eqnarray}
c(x,t)=c_{0} \exp(i\beta(x,t)), \quad
d(x,t)=d_{0} \exp(i\beta(x,t)).
\label{seq5}
\end{eqnarray} 
Using
$
\beta(x,t)=q x-\Omega t,
$
the dispersion relation connecting the  wavenumber $q$ and
frequency $\Omega$ of the perturbation (see e.g., \cite{sulem})
\begin{equation}
(-\Omega+2 k q)^{2}=q^{2} (q^{2}-2 U \psi_0^{2})
\label{seq6}
\end{equation}
is obtained.
This implies that the instability region for Eq. (\ref{seq1})
appears for perturbation wavenumbers 
$q^2<{2 U} \psi_0^2$, 
and in particular {\it only for focusing nonlinearities}
(to which we restrict this study).

We now attempt to identify the interval of unstable wavenumbers
by means of the TDVA. In particular, we start from the
Lagrangian L
\begin{eqnarray}
L=\int_{-\infty}^{\infty} \left[ \frac{i}{2} \left(\psi^{\ast} \psi_t-
\psi \psi_t^{\ast} \right) - 
|\psi_x|^2+ \frac{U}{2} |\psi|^4 \right]~ dx,
\label{seq7}
\end{eqnarray}
and consider a modulation of the plane wave of the form
\begin{eqnarray}
\psi=[\psi_0+a(t) \exp(i \phi_a(t)) \exp(i q x)+b(t) \exp(i \phi_b(t)) 
\exp(-i q x)] \exp\left(i(k x-\omega t) \right).
\label{seq8}
\end{eqnarray}
However, instead of considering the modulation directly at the level
of the equation, the variation of our approach is that 
{\it we use the modulational ansatz in the Lagrangian}. This constitutes
the basic novel ingredient of this variational-type approach to the 
modulational instability.

Here we consider an annular (1-dimensional) geometry, which imposes
periodic boundary conditions on the wavefunction $\psi(x)$ and 
integration limits
$0 \leq x < 2 \pi$ in Eq. (\ref{seq7}). This results in the quantization
of the wavenumbers $k,q = 0, \pm 1, \pm 2,...$.
However, it is clear that our results can be easily generalized to the 
case of an infinite, open system and to higher dimensions.

After substitution of Eq. (\ref{seq8}) into Eq. (\ref{seq7}),
we obtain the variational Lagrangian
\begin{eqnarray}
L=&&\pi [-2 (a^2 \dot{\phi}_a+b^2 \dot{\phi}_b)+
2 (U \psi_0^2 -q^2)(a^2+ b^2)-U \psi_0^4-4 q k (a^2-b^2)
+4 U \psi_0^2 a b \cos(\phi_a +\phi_b) \cr
&&+U (a^4+b^4+4 a^2 b^2)].
\label{seq9}
\end{eqnarray}
It is clear from this Lagrangian that the pair
$\phi_a(t), \phi_b(t)$ can be interpreted
as the generalized coordinates of the system, 
while $A(t) = 2 a^2(t), B(t) = 2 b^2(t)$ are
the corresponding momenta. 
In particular, the pairs $A(t),\phi_a(t)$ and $B(t),\phi_b(t)$
are canonically conjugate with respect to the effective Hamiltonian 
\begin{eqnarray}
H_{eff} = (U \psi_0^2 -q^2)(A + B)-2 q k (A - B)+
2 U \psi_0^2 \sqrt{AB} \cos(\phi_a +\phi_b) 
+{U \over 4}  (A^2+B^2+4 A B),
\label{ham}
\end{eqnarray} 
which is an exact integral of motion on the subspace spanned by 
Eq. (\ref{seq8}).

The Lagrangian equations of motion are: 
\begin{eqnarray}
\frac{d}{dt} \frac{\partial L}{\partial \dot{a}} &=&
\frac{\partial L}{\partial a} \Rightarrow
a \dot{\phi}_a= C_1 a+C_2 b \cos(\phi_a+\phi_b)+ U a (a^2+2 b^2)
\label{seq10}
\\
\frac{d}{dt} \frac{\partial L}{\partial \dot{\phi}_a} &=& 
\frac{\partial L}{\partial \phi_a} \Rightarrow 
\dot{a}=C_2 b \sin(\phi_a+\phi_b)
\label{seq11}
\\
\frac{d}{dt} \frac{\partial L}{\partial \dot{b}} &=&
\frac{\partial L}{\partial b} \Rightarrow
b \dot{\phi}_b= C_3 b+C_2 a \cos(\phi_a+\phi_b)+ U b (b^2+2 a^2)
\label{seq12}
\\
\frac{d}{dt} \frac{\partial L}{\partial \dot{\phi}_b} &=& 
\frac{\partial L}{\partial \phi_b} \Rightarrow 
\dot{b}=C_2 a \sin(\phi_a+\phi_b),
\label{seq13}
\end{eqnarray}
where $C_1 = U \psi_0^2-q^2-2 q k$, $C_2 = U \psi_0^2$,
and $C_3 = U \psi_0^2-q^2+2 q k$ are constant prefactors.

If we now keep all terms to O$(a)$ in 
Eqs. (\ref{seq10})-(\ref{seq13}) [which 
is consistent with an approximation linear
in $a$], we obtain
\begin{eqnarray}
a&=&b
\label{seq17}
\\
\dot{a}&=&C_2 a \sin(\phi)
\label{seq18}
\\
\dot{\phi}&=&(C_1+C_3)+2 C_2 \cos(\phi),
\label{seq19}
\end{eqnarray}

where $\phi=\phi_a+\phi_b$. 
The latter equation has the solution
\begin{eqnarray}
\phi(t)=2 \arctan\left[\frac{\sqrt{(2 U \psi_0^2-q^2)q^2}}{q^2} \tanh
\left(\sqrt{(2 U \psi_0^2-q^2)q^2} t \right)\right].
%\phi(t)=2 \arctan\left[\frac{\sqrt((2 C_3)^2-(C_1+C_2)^2)}{2 C_3-C_1-C_2} 
%tanh(\sqrt((2 C_3)^2-(C_1+C_2)^2) t)
%\right]
\label{seq20}
\end{eqnarray}

Two different cases arise here, corresponding respectively to whether
the instability criterion is satisfied or not. Namely,
when $2 U \psi_0^2-q^2<0$ the solution of Eq. (\ref{seq18}) is
\begin{eqnarray} 
a(t) \sim \sqrt {1-\frac{2 U \psi_0^2}{q^2} \sin 
\left(\sqrt{(q^2-2 U \psi_0^2)q^2} 
 t \right)^2},
\label{seq21}
\end{eqnarray}

while when $2 U \psi_0^2-q^2>0$ the solution of Eq.(\ref{seq18}) is
\begin{eqnarray} 
a(t) \sim \sqrt{1+\frac{2 U \psi_0^2}{q^2} \sinh
\left(\sqrt{(2 U \psi_0^2-q^2)q^2}
 t \right)^2}.
\label{seq22}
\end{eqnarray}

\begin{figure}
\centering
{\epsfig{file=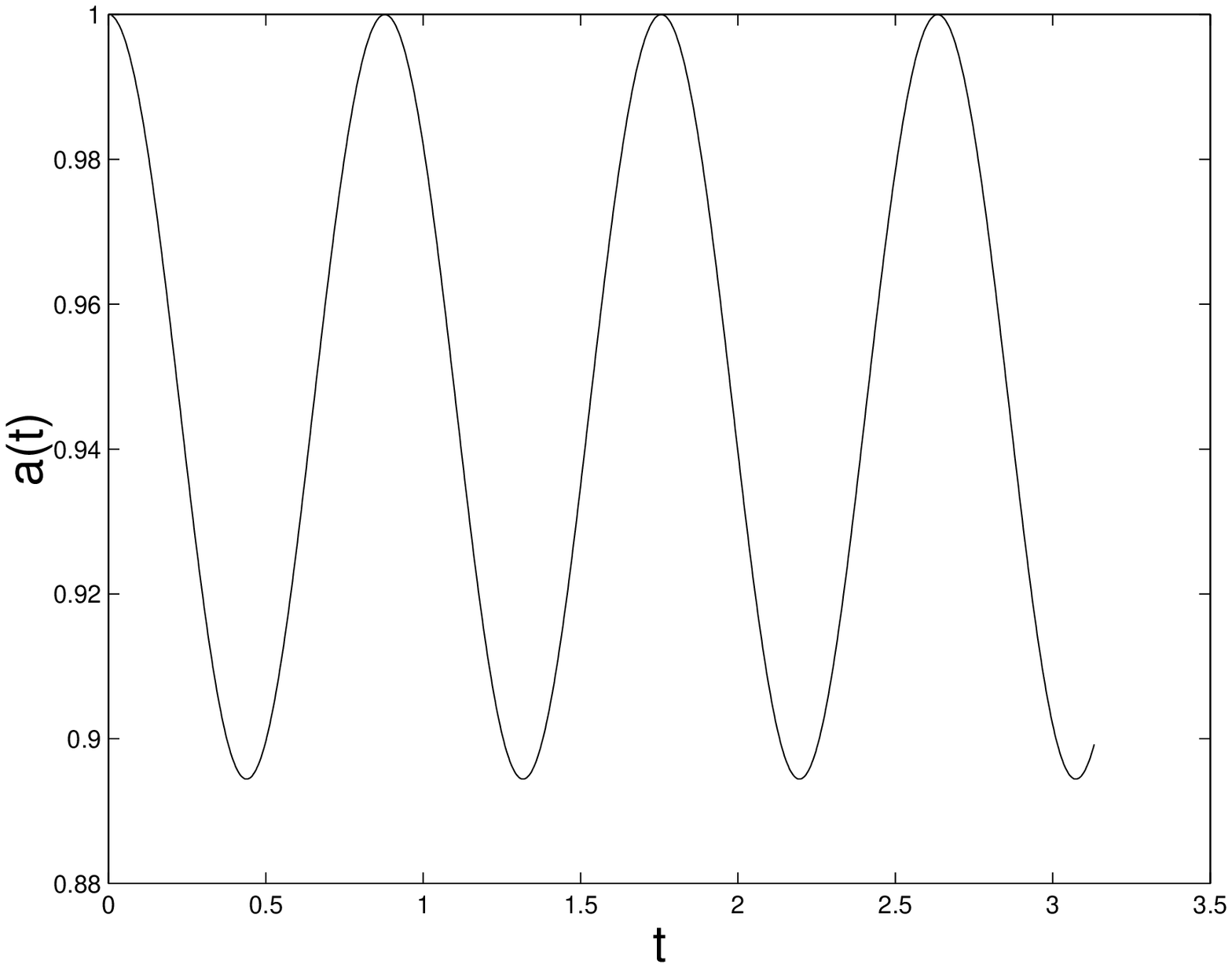, width=6.3cm,angle=0, clip=}}
{\epsfig{file=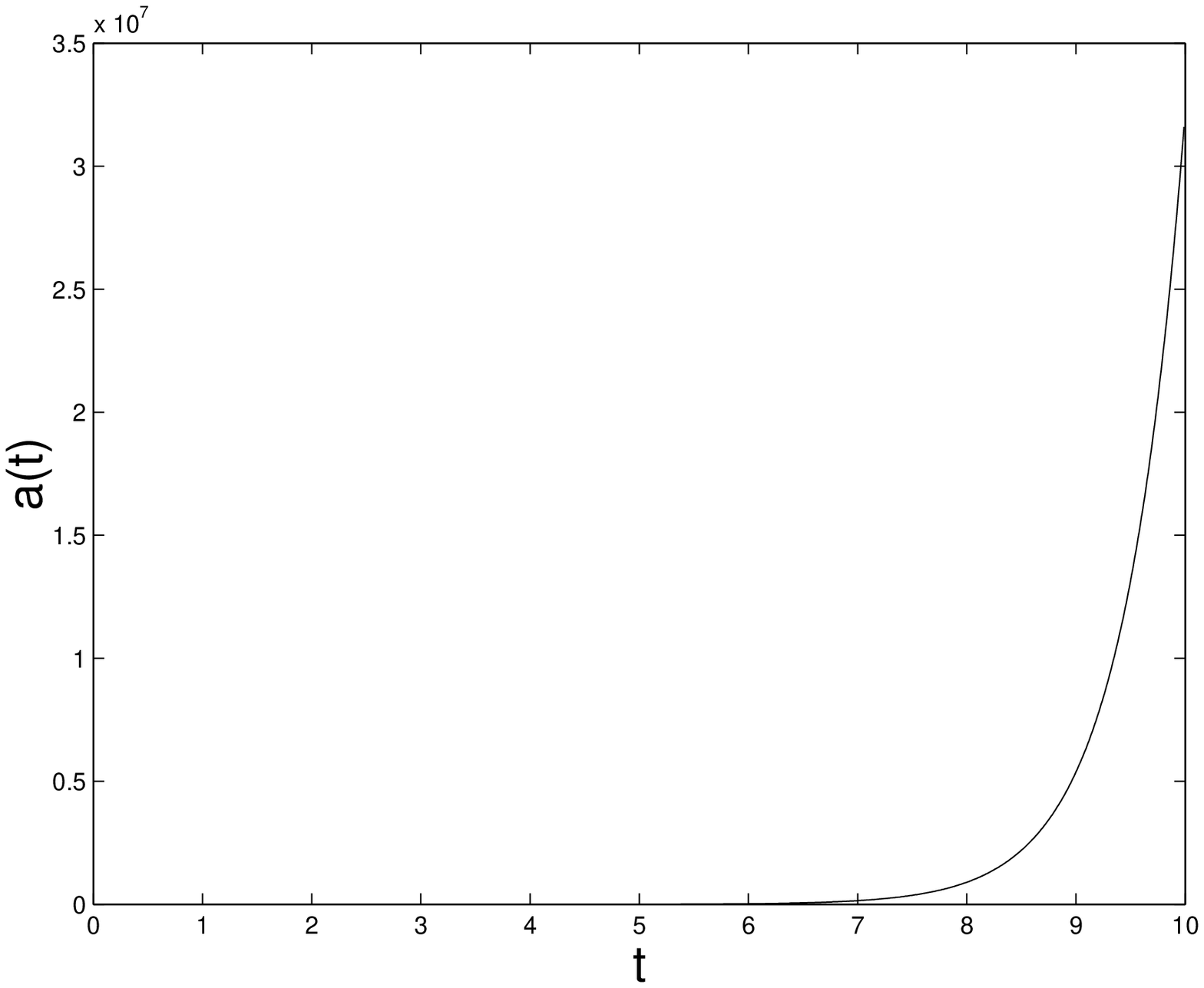, width=6.3cm,angle=0, clip=}}
\caption{The left panel shows the (stable oscillatory) 
time evolution of $a(t)$ for the case  $\frac{2 U \psi_0^2}
{q^2}=0.2$, in accordance with Eq. (\ref{seq21}). 
%Stable, oscillatory
%evolution.
The right panel shows the (unstable, exponentially growing) time evolution 
of $a(t)$ for the case of $\frac{2 U \psi_0^2}
{q^2}=1.2$, in accordance with Eq. (\ref{seq22}).}
\label{sfig1}
\end{figure}

The solutions signal the appearance of the modulational instability
when the threshold condition $q^2=2 U \psi_0^2$ is crossed (passing
from higher to lower perturbation wavenumbers).
This is also clearly shown in the time evolution of $a(t)$ in
accordance with Eqs. (\ref{seq21})-(\ref{seq22}) also shown in
Fig. \ref{sfig1} in the case of 
$q=2$ for $\frac{2 U \psi_0^2}
{q^2}=0.2$ (see the left panel of 
Fig. \ref{sfig1})  and $\frac{2 U \psi_0^2}{q^2}=1.2 $ 
(see the right panel of Fig. \ref{sfig1}).

%\begin{figure}
%\centering
%{\epsfig{file=graph1.ps, width=6.3cm,angle=0, clip=}}
%\caption{The time evolution of $a(t)$ for the case of $\frac{2 U \psi_0^2}
%{q^2}=1.2$, in accordance with Eq. (\ref{seq21}). Unstable, exponentially
%growing evolution.}
%\label{sfig2}
%\end{figure}

\begin{figure}
\centering
{\epsfig{file=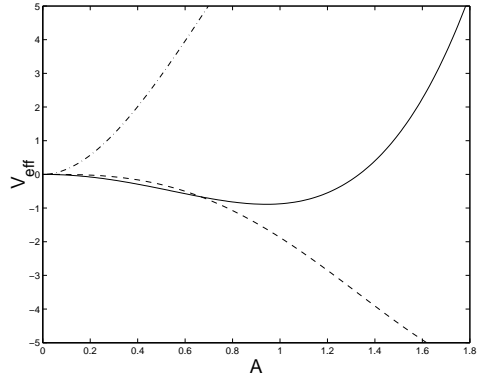, width=6.3cm,angle=0, clip=}}
\caption{The effective potential of Eq. (\ref{hamnew2}) is shown for
as a function of $A$ for $U=\psi_0=1$ and three different values
of $q$: $q=1$ (modulationally unstable; solid line), $q=\sqrt{2}$ 
(at the threshold; dashed line) and $q=2$ (modulationally stable; dash-dotted
line).}
\label{sfig3}
\end{figure}

An alternative, more intuitive way to appreciate the linear stability
result from a dynamical systems viewpoint. This consists of reducing
Eqs. (\ref{seq10})-(\ref{seq13}) to a one degree of freedom setting
with an effective potential energy landscape whose (parametric) variation 
will elucidate the instability. Along these lines, using $A(t=0)=0$ in
(without loss of generality) in Eq. (\ref{seq9}) and 
$A(t)=B(t)$ (from Eqs. (\ref{seq11}) and (\ref{seq13})) in Eq. (\ref{ham}),
we have:
\begin{eqnarray}
\tilde{H}= \pi \left[2 (U \psi_0^2-q^2)A+\frac{3}{2} U A^2+
2 U \psi_0^2 A \cos{\phi}\right] = 0.
\label{hamnew}
\end{eqnarray} 
Eliminitating $\phi$ from  Eqs. (\ref{seq11}) and (\ref{hamnew})
for $A(0)=0$,
we obtain the ``energy equation'' for $A$
\begin{eqnarray}
\frac{1}{2} \dot{A}^2 + V_{eff}=0,
\label{hamnew1}
\end{eqnarray}
where the effective potential $V_{eff}(A)$ is of the form:
\begin{eqnarray}
V_{eff}(A)=2 q^2(q^2-2 U \psi_0^2)A^2+ 3 U(U \psi_0^2-q^2) A^3
+\frac{9}{8} U^2 A^4
\label{hamnew2}
\end{eqnarray}
One can then examine the stability of the effective potential by 
evaluating its curvature at $A=0$. We thus obtain 
$V_{eff}''(A)|_{A=0}=4 q^2 (q^2-2 U \psi_0^2)$ and 
hence the
potential will be convex (and therefore the dynamics will be stable)
for $q^2 > 2 U \psi_0^2$, while it will be concave (and the dynamics
unstable) for $q^2 < 2 U \psi_0^2$. Hence in this case also, 
we retrieve the modulational stability criterion. The
effective potential is shown for the modulationally stable, unstable
and marginal case in Fig. \ref{sfig3}.

Now we turn to a more interesting case, where the coefficient
of the dispersion term, as well as the coefficient of the nonlinear
term in Eq. (\ref{seq1}) are temporally modulated, namely we examine
the equation
\begin{eqnarray}
i \psi_t =- D(t) \psi_{xx} - U(t) |\psi|^2 \psi.
\label{seq26b}
\end{eqnarray}
Our aim is to derive the modulational stability equation via the TDVA, for 
general $D(t)$ and $U(t)$. It is interesting to note that this equation has
become of increasing importance in the past decade due to applications both
in optics and also, more recently, in soft condensed-matter physics. In 
particular, in optics, the case of $D(t)$ periodic and $U(t)$ constant
is of relevance in the context of the so-called dispersion management. 
The latter is based on periodic alternation of fibers
with opposite signs of the group-velocity dispersion \cite{DM}. We note
in passing that in this application time $t$ is, in reality, the propagation 
distance (i.e., space), while $x$ corresponds to a retarded time variable.
An alternative setting where $D(t)$ is constant, but $U(t)$ can be 
temporally modulated (via a Feshbach resonance, i.e., an external magnetic
field; see e.g., \cite{inouye}) can be found in Bose-Einstein condensation.
In the latter setting, there has been an explosion of interest recently
in time dependent scattering length and its effect on patterns,
coherent structures and collapse thereof (a number of very recent
references can be found in \cite{frm,frm1,frm0,frm2}. 
While our primary motivation in considering 
MI through the TDVA in Eq. (\ref{seq26b}) principally stems from this
recently explored experimental potential in Bose-Einstein condensates,
we should note that this type of problem was investigated earlier 
in nonlinear optics, see e.g., \cite{jared}.

We consider the perturbation of the form
\begin{eqnarray}
\psi=\psi_{1} \left[1 + w(t) \cos(q x) \right]
\label{seq26c}
\end{eqnarray}
to the plane wave solution $\psi_1=e^{i (-k^2 \int_0^t D(s) ds+ \int_0^t U(s) ds+k x)}$.
Notice that in Eq. (\ref{seq26c}), we are using for simplicity
a variant of the ansatz of Eq. (\ref{seq8}), with $a=b$, $\phi_a=\phi_b$ and (without loss of generality) $\psi_0=1$.
Following the same procedure as above, we can obtain the stability
equations for $w=w_r + i w_i$:
\begin{eqnarray}
\dot{w}_r= q^2 D(t) w_i - \frac{3}{4} U(t) \left( w_i^3 + w_i w_r^2 \right)
\label{seq26d}
\\
\dot{w}_i= - \left[q^2 D(t)-2 U(t) \right] w_r + 
\frac{3}{4} U(t) \left(w_r^3 + w_r w_i^2 \right)
\label{seq26e}
\end{eqnarray}
Hence, at the linear level, we can derive the following stability 
equation
\begin{eqnarray}
\ddot{w}_r= \frac{\dot{D}(t)}{D(t)} \dot{w}_r -q^2 D(t)  
\left[q^2 D(t)-2 U(t) \right] w_r          
\label{seq26f}
\end{eqnarray}
By determining the windows of stability of the ordinary differential 
equation of (\ref{seq26f}), the modulational stability of Eq. (\ref{seq26b})
is determined. It is further worth noting that for $D(t)$ constant
and $U(t)$ time-periodic, Eq. (\ref{seq26f}) falls into Eq. (2) of
\cite{frm2} and becomes Hill's equation for which many stability results
are known in the mathematical literature \cite{magnus}. Furthermore,
in the case of $U(t)=1 + 2 \alpha \cos(\omega t)$, Eq. (\ref{seq26f}) 
falls into Eq. (2) of \cite{frm} and the resulting equation is of the
Mathieu type for which explicit stability windows can be computed
(for details see \cite{frm} and references therein). It then becomes
naturally an interesting problem in mathematical physics to determine
the stability of Eq. (\ref{seq26f}) for more general cases (e.g.,
with both coefficients periodically varying etc.).

%Eq. (\ref{seq23}) shows that we can study the modulational stability of 
%the problem in terms of a simple mechanical analog, namely interpreting 
%$\phi$ as the coordinate of a {\it classical} particle moving in the 
% potential $V(\phi)$. Notice that for simplicity, we are using a variant
%of the ansatz of Eq.

In this brief report, we have revisited the modulational instability
from a different point of view, namely a variational one. We have used
this dynamical systems' type approach to derive the Euler-Lagrange equations
for the time-dependent perturbation ansatz parameters and have examined
their stability for different wavenumbers of the perturbation (which affect
the constants of the ensuing set of ordinary differential equations). 
We have retrieved, in a simple and intuitive way, the criterion for the 
instability. The technique has also been generalized in cases in which
the coefficients of the dispersion and/or nonlinearity are temporally
varying (a case which we have argued to be relevant to a variety of 
applications). We have found the corresponding stability condition
obtaining a novel ordinary differential equation, whose special cases
correspond to stability/instability criteria established previously.
It would be interesting to extend the considerations of
this method (which seems applicable to any setting with an underlying 
Lagrangian/Hamiltonian structure) to contexts with explicit spatial
dependence of the potential (see e.g., \cite{sm02,cat02}).

The support of
NSF (DMS-0204585), UMass and the Clay Institute (PGK) is gratefully
acknowledged. Work at Los Alamos is supported by the US DoE.

%\end{multicols}{2}


\begin{thebibliography}{99}

\bibitem{sulem} C. Sulem and P.L. Sulem,
\newblock {\it The Nonlinear Schr{\"o}dinger Equation},
Springer-Verlag (New York, 1999).


\bibitem{benjamin67} T.B. Benjamin and J.E. Feir, 
J. Fluid. Mech. {\bf 27}, 417 (1967).

\bibitem{ostrovskii69} L.A. Ostrovskii, 
Sov. Phys. JETP {\bf 24}, 797 (1969).

\bibitem{taniuti68} T. Taniuti and H. Washimi, 
Phys. Rev. Lett. {\bf 21}, 209 (1968); A. Hasegawa, 
Phys. Rev. Lett. {\bf 24}, 1165 (1970).

\bibitem{peyrard93} M. Peyrard, T. Dauxois, H. Hoyet and C.R. Willis, 
Physica {\bf 68D}, 104 (1993).

\bibitem{morandotti99}  R. Morandotti, U. Peschel, J.S. Aitchison,
\newblock H.S. Eisenberg and Y. Silberberg,
Phys. Rev. Lett. {\bf 83}, 2726 (1999).
%;
%H. Eisenberg, Y. Silberberg, R. Morandotti,
%\newblock A.R. Boyd and J.S. Aitchison,
%\newblock { Phys. Rev. Lett.} {\bf 81}, 3383-3386 (1998). 

\bibitem{kivshar92}  Yu.S. Kivshar and M. Peyrard, 
Phys. Rev. A {\bf 46}, 3198 (1992).

\bibitem{wu01} B. Wu and Q. Niu,
Phys. Rev. A {\bf 64}, 061603(R) (2001).

\bibitem{sm02} A. Smerzi, A. Trombettoni, P.G. Kevrekidis and 
A.R. Bishop,
%Superfluid-Insulator Transition in a Chain of Weakly
%Coupled Bose-Einstein Condensates, 
Phys. Rev. Lett., {\bf 89}, 170402 (2002).

\bibitem{konotop02}  V.~V.~Konotop, and M.~Salerno, \pra {\bf
65}, 021602 (2002).

\bibitem{cat02} F. S. Cataliotti {\it et al.},
New Journal of Physics 5, 71 (2003). 

\bibitem{malomed} B.A. Malomed, Prog. Opt. {\bf 43}, 71 (2002).

\bibitem{treppiedi} B. Crosignani, P. Di Port and A. Treppiedi,
\newblock Quantum Semiclass. Opt. {\bf 7}, 73 (1995).

\bibitem{pra} T.A.B. Kennedy, Phys. Rev. A {\bf 44}, 2113 (1991);
S.J. Carter, P.D. Drummond, M.D. Reid and R.M. Shelby,
Phys. Rev. Lett. {\bf 58}, 1841 (1987);
P.D. Drummond, S.J. Carter and R.M. Shelby,
\newblock Opt. Lett. {\bf 14}, 373 (1989);
H. Haus and Y. Lai, J. Opt. Soc. Am. B
{\bf 7}, 386 (1990); 
H.P. Thacker, Rev. Mod. Phys. {\bf 53}, 253 
(1981).

\bibitem{trillo} S. Trillo and S. Wabnitz, Opt. Lett. {\bf 16},
986 (1991); G. Cappellini and S. Trillo, J. Opt. Soc. Am. B
{\bf 8}, 824 (1991); S. Trillo and S. Wabnitz, Opt. Lett.
{\bf 16}, 1566 (1991).



\bibitem{DM} C. Kurtzke, IEEE Photon. Technol. Lett. {\bf 5}, 1250 (1993);
N.J. Smith {\it et al.}, Electron. Lett. {\bf 32}, 54 (1996);
I. Gabitov and S. Turitsyn, Opt. Lett. {\bf 21}, 327 (1996);
N.J. Smith and N.J. Doran, Opt. Lett. {\bf 21}, 570 (1996).

\bibitem{inouye}  S. Inouye {\it et al.}, Nature {\bf 392}, 151 (1998); 
J. Stenger {\it et al.}, Phys. Rev. Lett. {\bf 82}, 2422 (1999).

\bibitem{frm} K. Staliunas, S. Longhi and G. J. de Valc\'{a}rcel, 
\newblock Phys. Rev. Lett. {\bf 89}, 210406 (2002).

\bibitem{frm1} F.Kh. Abdullaev, E.N. Tsoy, B.A. Malomed, R.A. Kraenkel,
cond-mat/0306281; F.Kh. Abdullaev, J.G. Caputo, R.A. Kraenkel, B.A. Malomed,
cond-mat/0209219. 

\bibitem{frm0} E.A. Donley {\it et al.}, Nature {\bf 412}, 295 (2001).

\bibitem{frm2} P.G. Kevrekidis, G. Theocharis, D.J. Frantzeskakis and
B.A. Malomed, Phys. Rev. Lett. {\bf 90}, 230401 (2003).

\bibitem{jared} J. Bronski and N. Kutz, Opt.Lett. {\bf 21}, 937 (1996); 
F.K. Abdullaev, S.A. Darmanyan, A. Kobyakov and F. Lederer. 
Phys.Lett.A {\bf 220}, 213, (1996).

\bibitem{magnus} W. Magnus and S. Winkler,
{\it Hill's Equation} (Wiley, New York, 1966).

%\bibitem{berman} G. Berman, A. Smerzi and A.R. Bishop,
%Phys. Rev. Lett., {\bf 88}, 120402 (2002)

%\bibitem{inst} L.S. Schulman, 
%\newblock {\it Techniques and applications of path integration}, Wiley,
%(New York, 1981).


\end{thebibliography}
\end{document}